\documentclass[pra,showpacs,twocolumn,superscriptaddress]{revtex4}
\usepackage{graphicx}
\usepackage{epsfig}
\usepackage{dcolumn}
\usepackage{bm}

\newcommand{\ket}[1]{|#1\rangle}

\newcommand{\ketbra}[2]{| #1 \rangle \langle #2 |}

\begin{document}

\title{Viable entanglement detection of unknown mixed states in low dimensions}
\author{Thiago O. Maciel} 
\email{maciel@gmail.com}
\author{Reinaldo O. Vianna}
\email{reinaldo@fisica.ufmg.br}
\affiliation{Departamento de F\'{\i}sica - ICEx - Universidade Federal de Minas Gerais,
Av. Ant\^onio Carlos 6627 - Belo Horizonte - MG - Brazil - 31270-901.}

\date{\today}

\begin{abstract}
We explore procedures to detect entanglement of unknown mixed states,
which can be experimentally viable. The heart of
the method is a hierarchy of simple feasibility problems, which provides 
sufficient conditions to entanglement.
Our numerical investigations
indicate that the entanglement is detected with a cost which is much lower than full
state tomography. The procedure is applicable to 
both free and bound entanglement, and involves only  single copy measurements.

\end{abstract}

\maketitle




\section{Introduction}

Entanglement is in the heart of Quantum Information, and allows for
promising spectacular applications like Quantum Computation and
Unconditionally Secure Quantum Cryptography. Our understanding of entanglement
has largely grown in the last few years, but the experimental detection
is still a daunting challenge. A recent and thorough review of the subject
is provided by the Horodecki family \cite{Horodecki-review}.

Theoretically, the tool of choice to detect entanglement is an {\em Entanglement Witness} (EW)
\cite{Horodecki-witness}. It consists of a Hermitian operator ($W$)  with 
non-negative expectation values for
all the separable states, but which can have a negative expectation
value for an entangled state, in this case, the state is said to be
detected by the EW.  When it comes to experiments, EWs are not that good,
for each state has its own optimal witness. The construction of the optimal EW depends on the knowledge of the state (see \cite{FGSLB-prl}, for example). There does not exist an EW which detects all the entangled density operators acting on a given Hilbert space.
Nevertheless, an EW can detect many states on a certain region of the state space,
though it will be optimal just for a restricted family of states. Therefore,
when some information about the state is known, an EW can be implemented.
For examples of experimental implementations of EWs, consult the references
in \cite{Horodecki-review}.

Exploring collective measurements  to estimate nonlinear 
functionals of quantum states,  Walborn {\em et al.} \cite{Walborn} have experimentally
measured the concurrence of unknown pure two qubit states, using two copies
of the objective state. It has also been  extended to the estimation of the concurrence of  mixed states, and implemented
experimentally \cite{Buchleinter}. 
In the case of rank-2 two-qubit states,
it is also possible to measure the concurrence exactly by means of collective
measurements on four copies \cite{Yu}.

Here we investigate measurements on single
copies of unknown mixed quantum states. 
The method we propose is simple, 
and shows to be effective in low dimensions. 
We are advocating that using sophisticated mathematical
tools to characterize entanglement in a data post-processing
fashion, while keeping the experiment as simple as possible, is efficient.
Therefore it is an approach in the opposite direction of works 
like those in \cite{Walborn, Buchleinter}, for example, where
entanglement is directly measured in a very elaborate experiment.
We have performed numerical tests in systems of two qubits,
one qubit and one qutrit and  two qutrits. 
In the case of two qutrits, we have investigated both bound and free entanglement. 
Though  we discuss
just bipartite entanglement, the formalism can be straightforwardly 
applied to the multipartite case.
 The basic idea is to consider the state written in
an orthonormal basis, which can be thought of as a {\em generalized Bloch
representation}. Then the projections, which are the components of the 
{\em generalized Bloch vector},  are gradually measured.
 For each set
of measurements,  it is checked if there is enough information to infer 
entanglement. In the case of states with Negative Partial Transpose (NPT),
we check if there is no state with Positive Partial Transpose (PPT) compatible
with the measurements. In the general case, including entangled PPT states \cite{ppt}, we build an entanglement 
witness compatible with the measurements, using the method described in \cite{FGSLB-prl}.
In the spirit of data post-processing, we mention  
the recent techniques independently introduced by Eisert {\em et al.} 
\cite{Eisert} and G\"uhne {\em et al.}  \cite{Guhne}, that yield  bounds
to certain entanglement quantifiers, based on the measurement of non-optimal
EWs.
We note also that Badziag {\em et al.} \cite{Badziag} and Hassan {\em et al.} \cite{Hassan}
have  introduced interesting entanglement criteria based on the norm
of the generalized Bloch vector.
In the context of Quantum Key Distribution, Curty {\em et al.}  \cite{Curty}
have introduced a method to check the presence of entanglement by 
means of EWs built with  previous measured data.

In the next section, we introduce and illustrate our method. 
Section III discusses a possible choice of informationanlly complete set of observables, 
and illustrates how our method would
perform with projective measurements. In particular, that section
makes clear that, compared to two qubits and two qutrits, 
it is not obvious how to choose the minimal amount of (or {\em optimal}) measurements
in a $2\otimes3$ system. Section IV offers more
questions than answers. There we analyze the limitations of 
the method, studying three representative states, namely, one highly
entangled and tow very low entangled states. As presented in 
this work, our method offers a yes/no answer about the entanglement
of a unknown mixed state, but the calculations of section IV suggest
that a further development of it could yield a good quantitative
estimator of entanglement. Section V concludes the paper.

\section{Theory}

Given a state represented by the density operator $\rho$, we want to
check if it is entangled, without performing a full tomography. 
As matter of fact, we want to make the least possible number of assumptions   about the
state.
We will present a strategy based on acquisition of partial information about
the state,
followed by data (post-)processing in form of Semi-Definite Programs (SDP).
SDPs  can be efficiently solved \cite{sedumi},
 and have exact solutions.
 As the numerical tests will show, it is effective in 
low dimensions. We focus on bipartite states
in order to simplify the discussion, but the generalization 
of the formalism to
multipartite states is straightforward.

A state  $\rho$, acting on the Hilbert space
$\mathcal{H}_d = \mathcal{C}^{d_a} \times \mathcal{C}^{d_b}$,
where $d=d_a \times d_b$,
can be written as:
\begin{equation}
\rho=\sum_{i=0}^{d_a^2-1}\sum_{j=0}^{d_b^2-1} r_{ij}P_{ij},
\end{equation}
where   the $P_{ij}$  are observables forming
a complete basis in the Hilbert-Schmidt space, and
 $r_{ij}=Tr(\rho P_{ij}) \in\mathcal{R}$.

 One possible choice for these
observables is $P_{ij}=\sigma_i^{d_a}\otimes\sigma_j^{d_b}$, with
the $\sigma_i^{d_s}$ being $SU(d_s)$ matrices, i.e., generalizations of
the Pauli matrices, where $\sigma_0^{d_s}$ stands for the identity matrix, and
$r_{00}=1/d$.
In this case, the state can also be written with explicit {\em local} and
{\em non-local} parts, and we have an expression that can be thought of as 
a {\em generalized Bloch representation}, namely,
\begin{eqnarray}
\label{rho}
\rho = \frac{1}{d}({\bf I_{d_a}}\otimes {\bf I_{d_b}} +
 \vec{r}_a \cdot \vec{\sigma}^{d_a} \otimes {\bf I_{d_b}} +
 {\bf I_{d_a}} \otimes
  \vec{r}_b \cdot \vec{\sigma}^{d_b} +  \nonumber \\
\sum_{i=1}^{d_a^2-1}\sum_{j=1}^{d_b^2-1} t_{ij}\sigma_i^{d_a}\otimes\sigma_j^{d_b}),
\end{eqnarray}
where  ${\bf I_{d_s}}$ is the $d_s\times d_s$ 
identity matrix, $\vec{\sigma}^{d_s}$ are the matrices for $SU(d_s)$,
$\vec{r}_s$ $\in$ $\mathcal{R}^{d_s^2-1}$,  
and finally $t_{ij}\in\mathcal{R}$. Note that $\vec{r}_a$ and $\vec{r}_b$ are 
the local parameters, defining the reduced density matrices, namely:
\begin{equation}
\label{marginals}
\rho_a \equiv Tr_{b}\rho = \frac{1}{d_a}(\bf{I_a} + \vec{r}\cdot
 \vec{\sigma}^{d_a}),
\end{equation}
where $Tr_{b}$ is the partial trace on subsystem $b$, and an analogous
expression for $\rho_b$. 
The {\em non-local} parameters,
\begin{equation}
\label{T}
 t_{ij} = Tr(\rho\sigma^{d_a}_i\otimes\sigma^{d_b}_j) = \langle T_{ij}\rangle ,
\end{equation}
 form a real matrix $T$, and are responsible for the classical and
quantum correlations in $\rho$. Note that the parameters in
Eq. 1 or Eq. 2   must be real in order to $\rho$ to be Hermitian,
 but it does not guarantee its 
positivity.

We will introduce some procedures to check the entanglement of
an unknown state, based on partial information about it.
This partial information consists of the knowledge of  some of the $r_{ij}$ (viz. Eq. 1),
eventually enriched with some further characteristic of the state, as the fact
that it is NPT, or its marginals are known (viz. Eq. 3).

As it is well known, it is  harder to check the entanglement of 
a bound entangled state than a free entangled one. 
From the theoretical point of view, it is easy to know
if the latter are entangled, for they have negative partial
transpose, which is known as the Peres-Horodecki criterion 
\cite{Horodecki-witness, Peres}.
But if the state is PPT, we need an entanglement witness.
The known examples of bound entangled states show very low
entanglement, therefore they will be more difficult to be checked
experimentally.

Now we present our first procedure, which checks entanglement in 
NPT bipartite states. The method we propose can be  thought of as a 
way of checking the Peres-Horodecki criterion. Assuming the 
knowledge of $n$ ($n\leq d^2-1$) of the parameters $r_{k}\equiv r_{ij}$
($k=1,2,\ldots,d^2-1$) in Eq. 1,
we check the existence of a PPT state compatible with the available information.
The nonexistence of such a state witnesses the entanglement of the state of interest.
This can be done by means of the following very simple SDP:

\begin{center}
\em determine $\varrho$
\begin{equation}
\label{sdp1}
 subject\,\, to
\left\{
\begin{array}{l}
 \varrho \geq 0 \\
 Tr(\varrho)=1  \\  
 \varrho^{\Gamma} \geq 0 \\
  Tr(\varrho P_{k}) =  r_{k},\,\,\, k=1,2,...,n\, .

\end{array}
\right.
\end{equation}
\end{center}
This SDP  is a {\em feasibility} program.
$\varrho^{\Gamma}$ stands for the partial transpose
of $\varrho$.  When this
program is infeasible, i.e., when there is no PPT state compatible with the available data,
we are certain that the unknown $\rho$ is NPT  and, therefore, entangled.

It could happen that the state of interest has passed through some known decoherence
channel, which restricts the state's marginals to some known form. One example is
the Werner states \cite{Werner}, which correspond to {\em depolarized} states whose marginals are
maximally mixed. The program in Eq. \ref{sdp1} can be easily modified to include this additional
information, which corresponds to further constraints in the SDP, namely:

\begin{center}
\em determine $\varrho$
\begin{equation}
\label{sdp2}
 subject\,\, to
\left\{
\begin{array}{l}
 \varrho \geq 0 \\
 Tr(\varrho)=1  \\  
 \varrho^{\Gamma} \geq 0 \\
 \varrho_a = \rho_a \\ 
 \varrho_b = \rho_b \\ 
  Tr(\varrho P_{k}) =  r_{k},\,\,\, k=1,2,...,n\, .

\end{array}
\right.
\end{equation}
\end{center}

Programs in Eq. \ref{sdp1} and in Eq. \ref{sdp2}  determine the projection of the state of interest
in a certain hyperplane (in the Hilbert-Schimidt space), and  check the existence of the family of PPT states with the same projection.
 If there is no such  state, it means that
the measured state is NPT, and therefore entangled. This suffices to check
entanglement in spaces $2\times 2$ (qubit-qubit) and $2\times 3$ 
(qubit-qutrit) \cite{Horodecki-witness}. In larger spaces,  this approach still works for the 
NPT states, but it will not detect entangled PPT states. Now we introduce a procedure that, in principle,
can detect both free and bound entangled states.  

When the state of interest is in a space which allows for bound entanglement \cite{ppt}, 
we need an entanglement witness to check if it is separable or not. 
If we eliminate the constraint of positivity of the partial transpose in 
the programs of Eqs. \ref{sdp1} and \ref{sdp2},  namely , $ \varrho^{\Gamma} \geq 0$,
those programs return a state $\varrho$, which can be PPT or not, compatible with the available
data.  We then build an optimal  entanglement witness ($W_\varrho$) to $\varrho$, and use it to
estimate the entanglement of $\rho$, i.e., $Tr(W_\varrho \rho) \sim Tr(W_\varrho \varrho)$.
Remember that an EW is a Hermitian operator with non-negative expectation values
on separable states, but which can have a negative expectation value on an
entangled state, in this case, we say that the EW detects the entangled state.
The optimal EW of a state yields the most negative  expectation value, when compared
to any other EW of the same kind, therefore $Tr(W_\varrho \rho) > Tr(W_\rho \rho)$. 
Note that EWs can be chosen to correspond to 
different entanglement quantifiers \cite{FGSLB-pra}. The EWs in this work have the
constraint $Tr(W)=1$, and correspond to the random robustness \cite{Vidal, ROV}, 
which measures how resilient to white noise is the entanglement.

We need an {\em error bar} to our entanglement estimate given by $Tr(W_\varrho \varrho)$.
In order to do that, we rewrite Eq. 1 as:
\begin{equation}
\varrho=\sum_{k=1}^{n}r_k P_k + \sum_{j=n+1}^{d^2}r_jP_j.
\end{equation}
The first summation corresponds to the known data. Let's call
it $\varrho_{known}$. Of course, $\rho_{known}=\varrho_{known}$.
The second summation is yielded either by the program in Eq. \ref{sdp1}) or
Eq. \ref{sdp2}, and we call it $\varrho_{unknown}$. Now we can write our
entanglement estimate as:
\begin{equation}
Tr(W_\varrho \rho)= Tr(W_\varrho \varrho) \pm | Tr(W_\varrho \varrho_{unknown})|.
\end{equation}

The techniques we use to build the optimal EW are  based on SDPs, and are  described in 
\cite{FGSLB-prl}. 

In Fig. 1, we show how  the method performs for two qubits, one qubit and one qutrit, and two qutrits.
Each graph is built out of $10^4$ random NPT
 entangled states.
We plot the {\em efficacy of entanglement detection} (i.e., number of states detected as entangled divided
by $10^4$) in the sample of
states, against the number of measured non-local parameters (viz. Eq. 4). 
Every time the program in Eq. 5 or in Eq. 6 is infeasible for a 
given state, it means that the measured data were sufficient to
detect  entanglement. 
In the case of two qutrits, we also test the EW approach of Eq. 8. 
A state is considered successfully detected as entangled, when both
$ Tr(W_\varrho \varrho) + | Tr(W_\varrho \varrho_{unknown})| $  and
$ Tr(W_\varrho \varrho) - | Tr(W_\varrho \varrho_{unknown})| $ are
negative. About 70\% of the states are detected as entangled, with an
effort which is roughly half of a full state tomography. On one hand,
the less entangled
is the state, more information we need to infer its entanglement. Therefore, 
the graphs show 100\% efficacy only when all the tomographic parameters are 
measured. On the other hand, highly entangled states are detected with the
knowledge of only a few non-local parameters.

\begin{figure}
\includegraphics[scale=0.5]{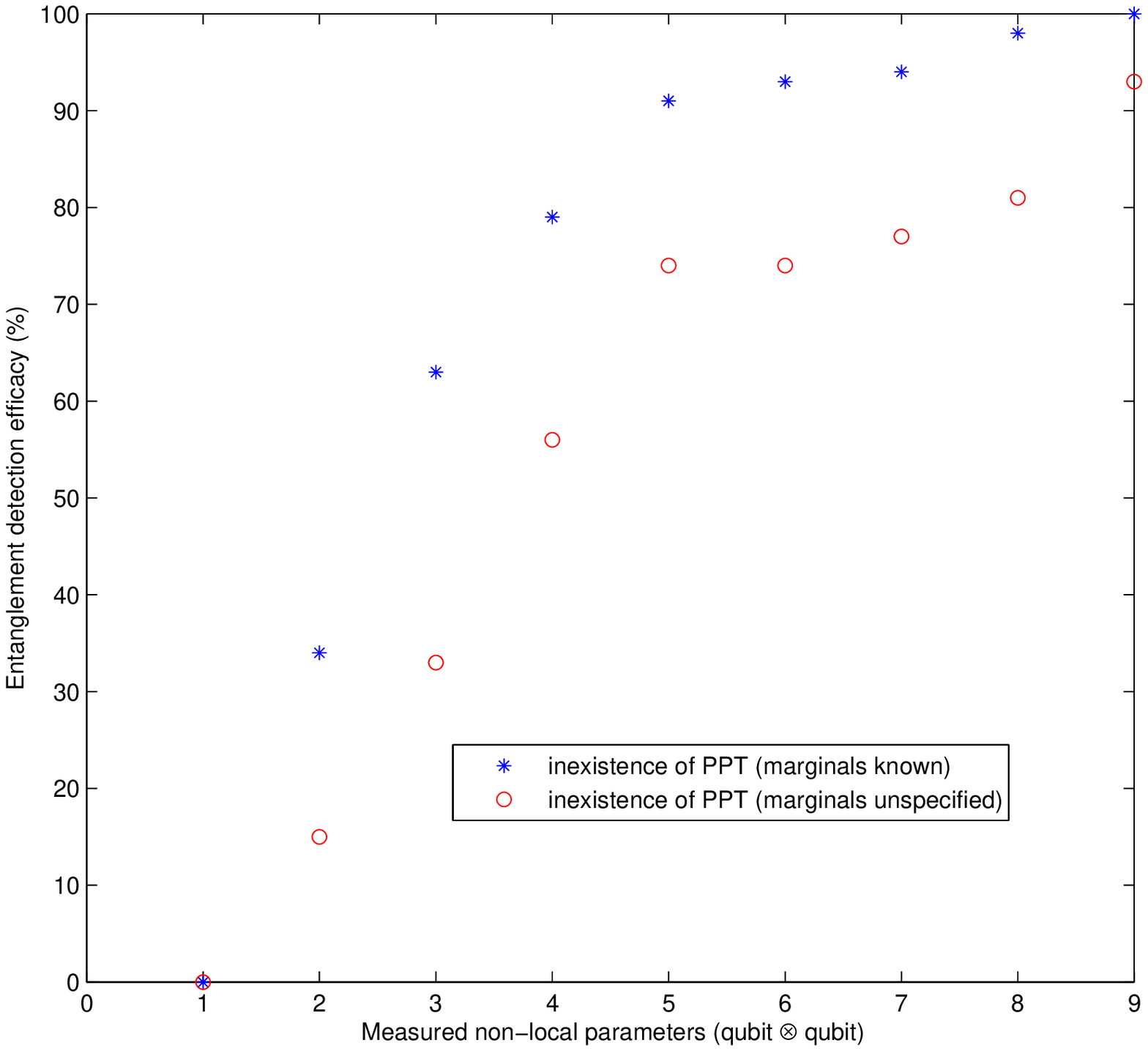}
\includegraphics[scale=0.5]{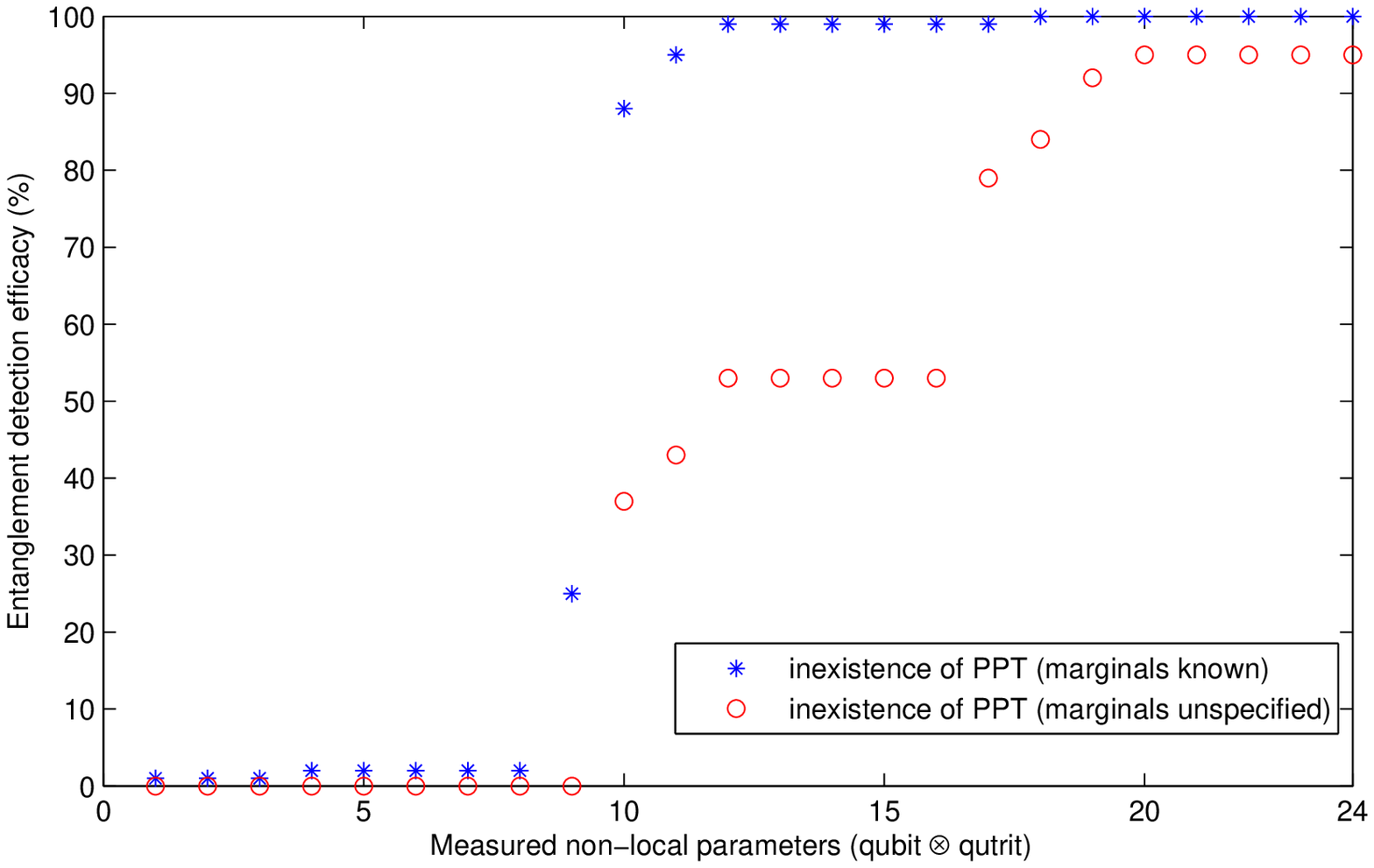}
\includegraphics[scale=0.5]{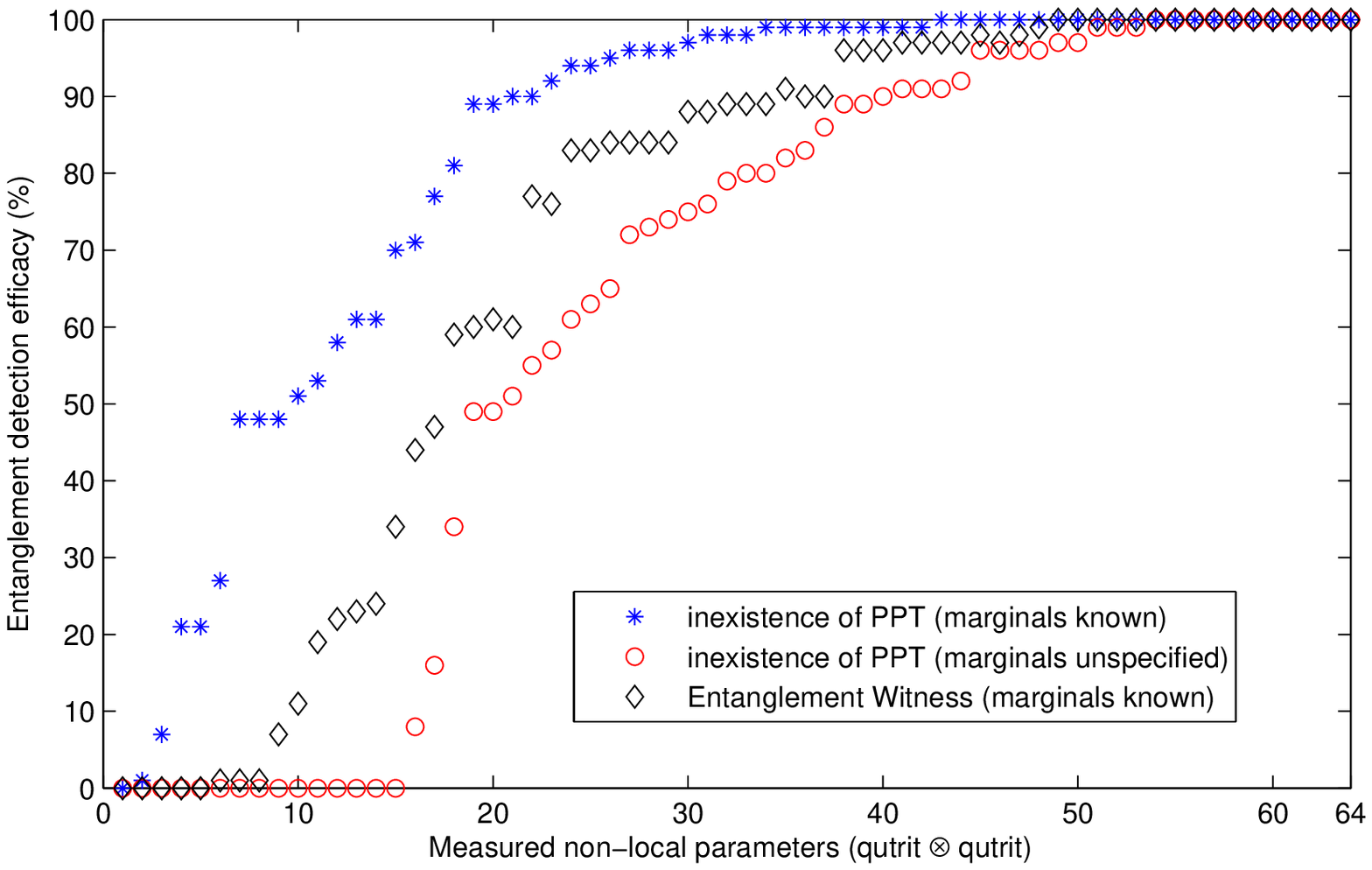}
\caption{Fraction of success of entanglement detection against
the number of measured non-local parameters (viz. Eq. 4) , for a sample of
$10^4$ random NPT states, using the approaches described in Eqs. 5 and 6, for
two qubits, one qubit and one qutrit, and two qutrits. For two qutrits, we
also show the results using the EW (Eq. 8).}
\end{figure}

\section{Choosing what to measure in the laboratory}

In the last section, we have described the general idea behind detecting
entanglement of unknown states, based on partial information. Our main goal
was to show that the proposed data post-processing is effective.
Now we want to discuss how  our technique could be actually implemented  with 
 projective measurements. A sensible way to do this is by grouping the observables
in the smallest number of maximal commuting classes. Commuting observables share
a common set of eigenvectors and, consequently, can be simultaneously measured (i.e., can be simultaneously diagonalized). Therefore,
such a scheme would yield the smallest number of complete projective measurements
to be done. 

Let's fix the basis of observables. For a Hilbert space of dimension $d_s$, we
introduce the {\em shift} and {\em clock} operators, namely:
\begin{equation}
X \equiv \sum_{j=0}^{d_s-1}\ketbra{j+1}{j} 
\end{equation}
and
\begin{equation}
Z \equiv \sum_{j=0}^{d_s-1}\exp(\frac{2\pi i j}{2})\ketbra{j}{j}, 
\end{equation}
where $\{\ket{j}, \,\, j=0,\ldots,d_s-1\}$ is an orthonormal basis. For dimension 2, these operators are the usual Pauli matrices. We also define:
\begin{equation}
Y \equiv  XZ
\end{equation}
and, for $d>2$,
\begin{equation}
V \equiv  XZ^2.
\end{equation}
Two-particle observables are now defined as tensor products of powers of these operators.
For two qubits, one qubit and one qutrit, and two qutrits, Table I
shows the complete bases of observables, with a convenient labelling 
\cite{Planat}.

\begin{table}
\begin{tabular}{c||c|c|c|c}
$2\otimes 2$ & I & Z & X & Y \\ \hline \hline 
I	  & 0 & 13 & 14 & 15 \\ \hline
Z	& 1 & 4 & 7 & 10 \\ \hline
X	& 2 & 5 & 8 & 11 \\ \hline
Y	& 3 & 6 & 9 & 12 \\ \hline \hline \hline
\end{tabular}

\begin{tabular}{c||c|c|c|c}
$2 \otimes 3$ & I & Z & X & Y \\ \hline \hline
I & 0 & 33 & 34 & 35 \\ \hline
Z & 1 & 9 & 17 & 25 \\ \hline
X & 2 & 10 & 18 & 26 \\ \hline
Y & 3 & 11 & 19 & 27 \\ \hline
V & 4 & 12 & 20 & 28 \\ \hline
$Z^2$ & 5 & 13 & 21 & 29 \\ \hline
$X^2$ & 6 & 14 & 22 & 30 \\ \hline
$Y^2$ & 7 & 15 & 23 & 31 \\ \hline
$V^2$ & 8 & 16 & 24 & 32 \\ \hline \hline \hline
\end{tabular}

\begin{tabular}{c||c|c|c|c|c|c|c|c|c}
$3 \otimes 3$ & I & Z & X & Y & V & $Z^2$ & $X^2$ & $Y^2$ & $V^2$ \\ \hline \hline
I & 0 & 73 & 74 & 75 & 76 & 77 & 78 & 79 & 80 \\ \hline
Z & 1 & 9 & 17 & 25 & 33 & 41 & 49 & 57 & 65 \\ \hline
X & 2 & 10 & 18 & 26 & 34 & 42 & 50 & 58 & 66 \\ \hline
Y & 3 & 11 & 19 & 27 & 35 & 43 & 51 & 59 & 67 \\ \hline
V & 4 & 12 & 20 & 28 & 36 & 44 & 52 & 60 & 68 \\ \hline
$Z^2$ & 5 & 13 & 21 & 29 & 37 & 45 & 53 & 61 & 69 \\ \hline
$X^2$ & 6 & 14 & 22 & 30 & 38 & 46 & 54 & 62 & 70 \\ \hline
$Y^2$ & 7 & 15 & 23 & 31 & 39 & 47 & 55 & 63 & 71 \\ \hline
$V^2$ & 8 & 16 & 24 & 32 & 40 & 48 & 56 & 64 & 72 \\ \hline
\end{tabular}

\caption{ Complete bases of observables in Hilbert spaces of dimensions
$2\otimes2$, $2\otimes3$ and $3\otimes3$. Each two-particle 
observable is the tensor product between one operator of the first line
by one operator of the first column.}
\end{table}	

Now we can group the observables of Table I in maximally commuting
classes. For two qubits, we have the 5 classes \cite{MUB} shown in
 Table II;
for one qubit and one qutrit,  there are 12 classes, as shown in Table III; and finally,
for two qutrits, there are the 10 classes \cite{Lawrence, Mean-King}
shown in Table IV.
Note that, for two qubits and two qutrits, the classes are disjoint sets. 
 In each case, the simultaneous eigenvectors of each class
form a set of Mutually Unbiased Bases (MUB) \cite{Ivanovic, Wootters}, in the sense that any 
two vectors of different bases have the same overlap's absolute value. 
In the case of the $2\otimes3$ system, the classes are neither disjoint and nor
minimal. With 5 observables in each class, the minimal number of classes, for 
a total of 35 distinct operators, 
should be 7. It is the case for the systems $2\otimes2$ and $3\otimes3$,
with 15 observables divided in 5 classes of 3 operators, and 80 observables
divided in 10 classes of 8 operators, respectively.
It is conjectured that there is no informationally  complete set (in the tomographic sense) of MUBs  for the $2\otimes3$ system 
\cite{Werner-HP, Englert, Klapp}, and it is known
that generalized Pauli matrices (which is our choice of observables) are not
extensible to MUBs \cite{Grassl}. 

\begin{table}
\begin{tabular}{llll}
$C_1=\{$ & 1 & 4 & 13 \} \\
$C_2=\{$ & 2 & 8 & 14 \} \\
$C_3=\{$ & 3 & 12 & 15 \} \\
$C_4=\{$ & 5 & 9 & 10 \} \\
$C_5=\{$ & 6 & 7 & 11 \} \\
\end{tabular}
\caption{Five maximally commuting classes of observables (viz. Table I) for two qubits.
The common eigenvectors of each class form a set of MUBs.}
\end{table}

\begin{table}
\begin{tabular}{rlllll}
$C_1=\{$ & 1 & 5 & 33 & 9 & 13 \} \\
$C_2=\{$ & 2 & 6 & 33 & 10 & 14 \} \\
$C_3=\{$ & 3 & 7 & 33 & 11 & 15 \} \\
$C_4=\{$ & 4 & 8 & 33 & 12 & 16 \} \\
$C_5=\{$ & 1 & 5 & 34 & 17 & 21 \} \\
$C_6=\{$ & 2 & 6 & 34 & 18 & 22 \} \\
$C_7=\{$ & 3 & 7 & 34 & 19 & 23 \} \\
$C_8=\{$ & 4 & 8 & 34 & 20 & 24 \} \\
$C_9=\{$ & 1 & 5 & 35 & 25 & 29 \} \\
$C_{10}=\{$ & 2 & 6 & 35 & 26 & 30 \} \\
$C_{11}=\{$ & 3 & 7 & 35 & 27 & 31 \} \\
$C_{12}=\{$ & 4 & 8 & 35 & 28 & 32 \} \\
\end{tabular}
\caption{Twelve maximally commuting classes of observables (viz. Table I) for qubit$\otimes$qutrit.}
\end{table}

\begin{table}
\begin{tabular}{rllllllll}
$C_1 =\{$ & 1 & 5 & 73 & 9 & 13 & 77 & 41 & 45 \} \\
$C_2 =\{$ & 2 & 6 & 74 & 18 & 22 & 78 & 50 & 54 \} \\
$C_3 =\{$ & 3 & 7 & 75 & 27 & 31 & 79 &  59 & 63 \} \\
$C_4 =\{$ & 4 & 8 & 76 & 36 & 40 & 80 & 68 & 72 \} \\
$C_5 =\{$ & 10 & 46 & 33 & 19 & 32 & 69 & 60 & 55 \} \\
$C_6 =\{$ & 11 & 47 & 17 & 28 & 38 & 53 & 66 & 64 \} \\
$C_7 =\{$ & 12 & 48 & 25 & 34 & 23 & 61 & 51 & 70 \} \\
$C_8 =\{$ & 14  & 42  & 29 & 20 & 39 & 57 & 67 & 56 \} \\
$C_9 =\{$ & 15  & 43  & 37 & 26 & 24 & 65 & 52 & 62 \} \\
$C_{10} =\{$ & 16  &  44 & 21 & 35 & 30 & 49 & 58 & 71 \} \\
\end{tabular}
\caption{Ten maximally commuting classes of observables (viz. Table I) for two qutrits.
The common eigenvectors of each class form a set of MUBs.}
\end{table}

In Fig. 2, we repeat the calculations of section II, namely Eqs. 1, 5 and 8,  but now using the MUB
projectors for the two qubits and two qutrits.

\begin{figure}
\includegraphics[scale=0.5]{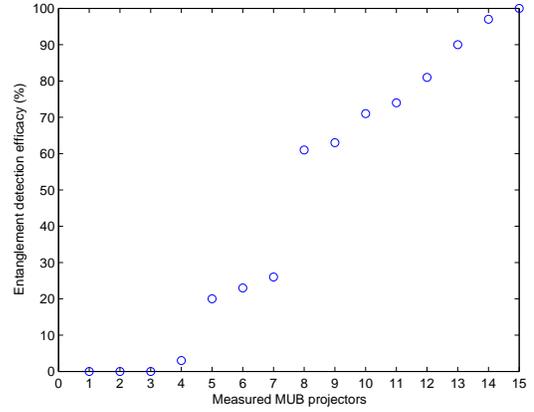}
\includegraphics[scale=0.5]{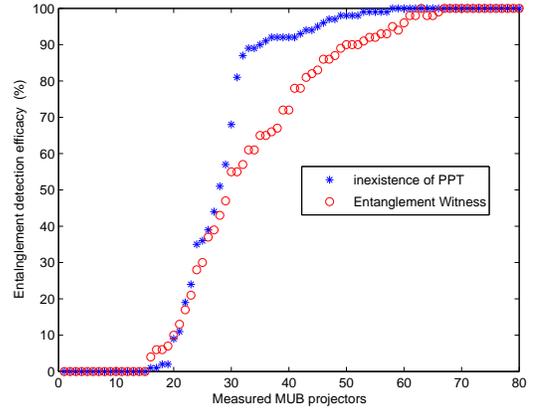}
\caption{Fraction of success of entanglement detection against
the number of measured MUB projectors  (viz. Eq. 1) , for a sample of
$10^4$ random NPT states, using the approaches described in Eq. 5 (Peres-Horodecki criterion), 
for two qubits (top), and Eqs. 5 and 8 (EW) for
two qutrits (bottom). }
\end{figure}

Though we do not know MUBs for the $2\otimes3$ system, we still want to
do the minimal number of projective measurements in the laboratory. To make a complete tomography,  we need a set of 35 informationally complete projectors.  
Measuring in the basis of common eigenvectors of each of the 12 classes
 (Table III), the numbers of independent projective measurements extracted
 from each class are, respectively, 5, 4, 4, 4, 3, 2, 2, 2, 3,  2, 2, 2.
These 35 projectors, which are linearly independent in the Hilbert-Schimidt
space, can be sorted in 7 sets of 5, and
re-orthonormalized  in order to correspond to 7 complete
projective measurements (7 {\em observables}). The results of measurements
in these two different bases are shown in Fig. 3.

\begin{figure}
\includegraphics[scale=0.5]{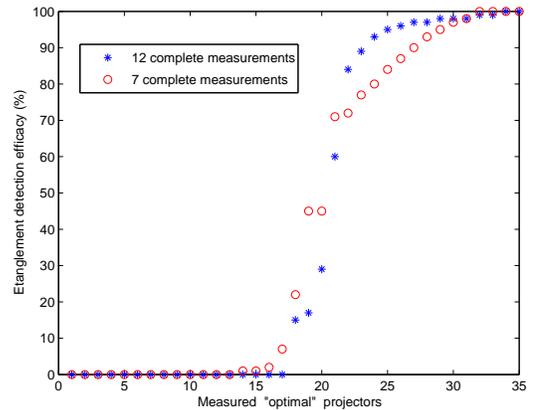}
\caption{Fraction of success of entanglement detection against
the number of {\em optimal}   projectors  (viz. Eq. 1 and Table III) , for a sample of
$10^4$ random NPT states, using the approach described in Eq. 5 (Peres-Horodecki criterion), for
the $2\otimes 3$ system. }
\end{figure}

\section{Estimated EW and Low-Entangled States}

In sections II and III, we have discussed how we could detect the entanglement
of unknown NPT states, based on partial information. In particular, the method
we have proposed to check the Peres-Horodecki criterion (Eq. 5) is 
rigorous, yields an exact answer and, according to our numerical tests, 
performs nicely for the 
systems $2\otimes2$, $2\otimes3$ and $3\otimes3$, as shown in Figs. 2 and 3.
On the other hand, our proposed EW estimate (Eq. 8) needs to be better understood.
Figs. 1 and 2 are numerical evidence for the correctness of Eq. 8, but we are lacking
a rigorous proof for the exact expression of our {\em error bar}.  In this section, we will study the performance of Eq. 8 
in  3 particular states, being one highly entangled two-qutrit Werner state \cite{Werner}
and two very low entangled states, being one of them also a two-qutrit Werner state, 
and the other one a two-qutrit bound entangled state \cite{Horodecki-bound}.
It will add further evidence of the correctness of Eq. 8, and will show that 
our proposed {\em error bar} is too big, i.e., it seems that $Tr(W_\varrho \varrho)$ is
a very good upper bound to $Tr(W_\varrho \rho)$, much better than we expected,
and there must be a tighter {\em error bar}, but we couldn't devise it yet. 
Note that $Tr(W_\varrho \rho)$ is certainly an upper bound to $Tr(W_\rho \rho)$.

\begin{figure}
\includegraphics[scale=0.5]{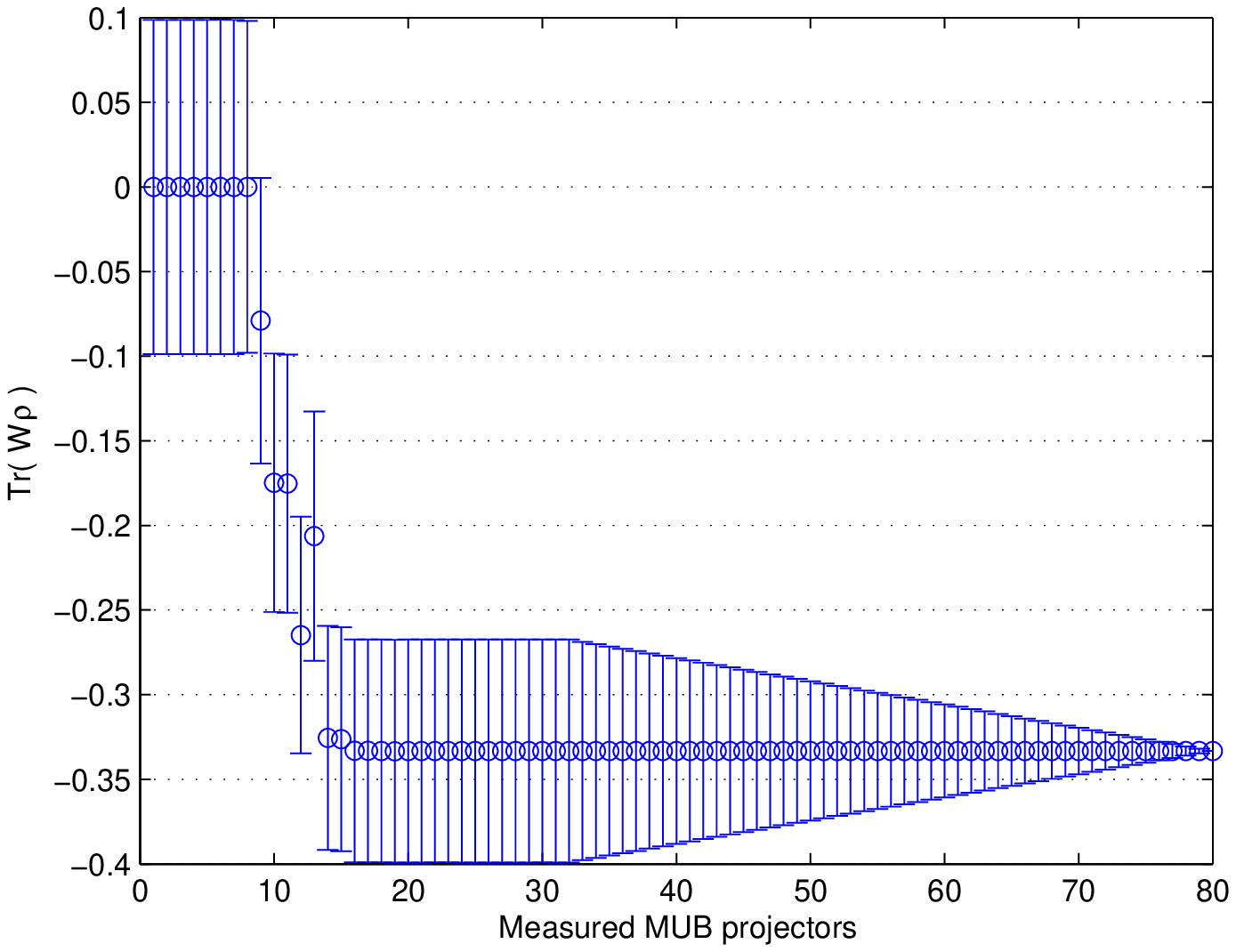}
\includegraphics[scale=0.5]{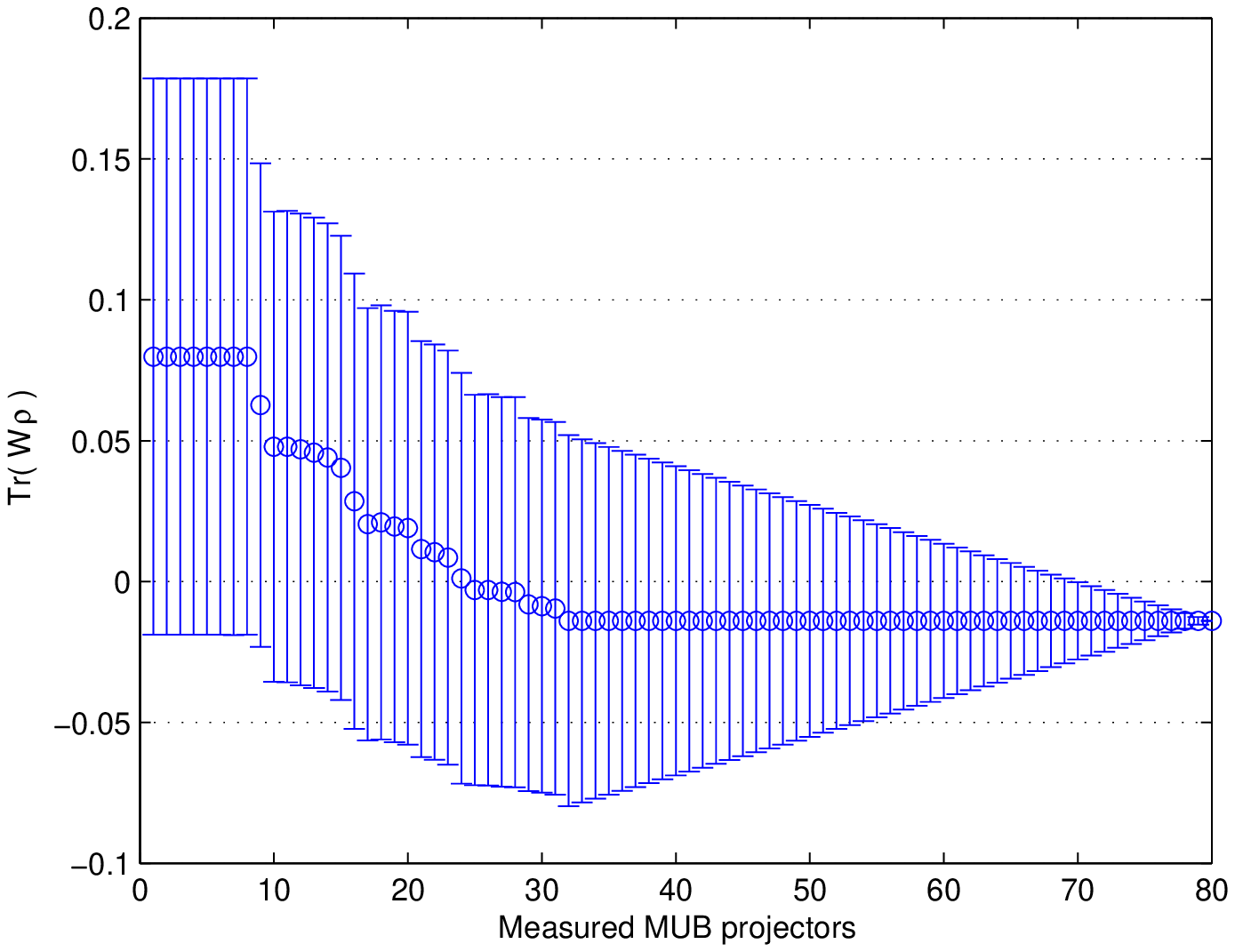}
\includegraphics[scale=0.5]{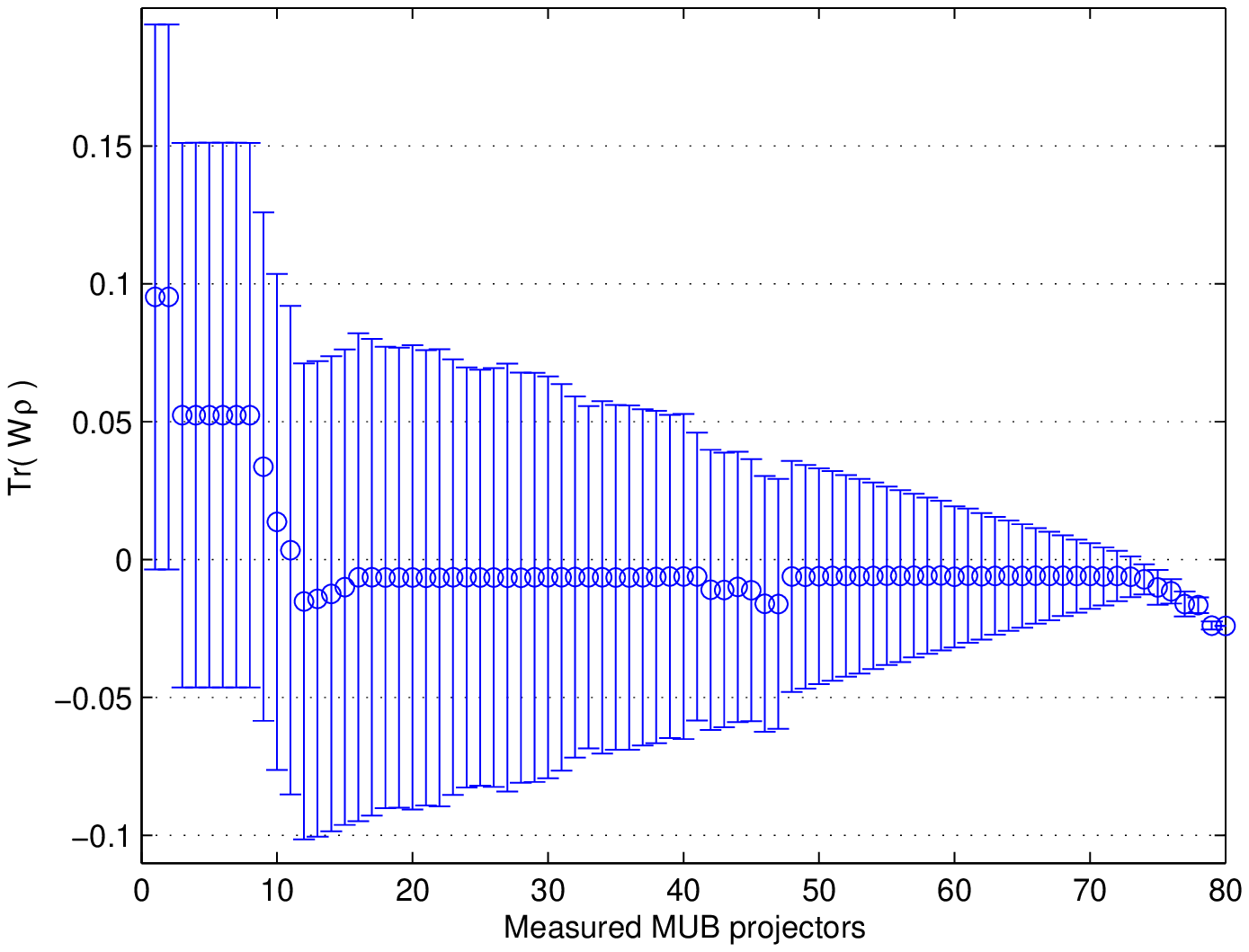}
\caption{Estimated EW with its {\em error bar} (Eq. 8) for 3 particular 
two-qutrit states: (top) Werner state with $\beta=-1$, (middle)
Werner state with $\beta=-0.37$, (bottom) Horodecki bound entangled state with
$\lambda=3.9$. In each case, the exact value for $Tr(W_\rho \rho)$ 
corresponds to the mark 80 in the abscissa.}
\end{figure}

The two-qudit (for our purposes $d=3$) Werner states \cite{Werner} can be written as:
\begin{equation}
\label{wern}
\rho_W=\frac{I_d+\beta  F_d}{d^2+d \beta},
\end{equation}
with $-1\leq \beta \leq 1$. $\rho_W$ is separable for
$\beta\geq-\frac{1}{d}$.
$F_d$ is a swap operator for two qudits,
\begin{equation}
 F_d= \sum_{i,j=1}^{d}\ketbra{ij}{ji} .
\end{equation}

The two-qutrit bound entangled state we use is picked up from the following family of states \cite{Horodecki-bound}:
\begin{equation}\label{rhosigma}
\rho_H  = \frac{2}{7}| \phi^+_3\rangle\langle \phi^+_3| + \frac{\lambda }{7} \sigma_+ + 
\frac{5- \lambda}{7} \sigma_- ,\,\,\, 2 \leq  \lambda \leq  5, 
\end{equation}
where
\begin{equation}\label{phiplus}
|\phi^+_3\rangle\langle \phi^+_3|  = \frac{1}{\sqrt{3}}\sum\limits_{i,j=0}^{2} |ii \rangle\langle jj|
\end{equation}
is the density matrix for the maximally entangled state, and
\begin{equation}\label{sigmaplus}
\sigma_+  = \frac{1}{3}(|01\rangle\langle 01|+ |12\rangle\langle 12| + |20\rangle\langle 20|),
\end{equation}
\begin{equation}\label{sigmaminus}
\sigma_-  = \frac{1}{3}(|10\rangle\langle 10|+ |21\rangle\langle 21| + |02\rangle\langle 02|)
\end{equation}
are two separable states. With these definitions, the character of $\rho_H$ changes with 
$\lambda$ according to
\begin{equation}\label{prop}
\rho_H= \left\{\begin{array}{ll}
separable, & 2 \leq  \lambda \leq  3, \\
bound\,\, entangled, & 3 <  \lambda \leq  4, \\
free\,\, entangled, &  4 <  \lambda \leq  5.
\end{array}\right. 
\end{equation}

In Fig. 4, we see the results yielded by Eq. 8 applied to Werner states (Eq. 13) with 
$\beta=-1$ ($Tr(W_{\rho_W} \rho_W)=-1/3$)  and $\beta=-0.37$ ($Tr(W_{\rho_W} \rho_W)=-0.014$), and to the Horodecki bound 
entangled state (Eq. 15) with $\lambda=3.9$ ($Tr(W_{\rho_H} \rho_H)=-0.024$). As we have mentioned before, a 
state is considered detected as entangled when the {\em error bar} resides entirely in the entangled region.
We see that the highly entangled Werner state is detected with just 11 measurements. On the other hand, the
two low entangled states are detected after the $70^{th}$ measurement. The Peres-Horodecki criterion (Eq. 5) 
detects the  low entangled Werner state in the $32^{nd}$ measurement but, of course, it is not applicable
to the PPT state. Ignoring the error bar, note that the estimated EW never super-estimated  the entanglement;
it  yielded the exact results for the Werner states after the $17^{th}$ measurement, for the most entangled
state, and after the $32^{nd}$ measurement, for the low entangled state; and, finally, it detected the 
bound entangled state after 10 measurements.

\section{Conclusion}

We discussed data post-processing strategies to characterize entanglement
of unknown mixed states, based on partial knowledge of the state.
The method is guaranteed to work, for it {\em converges} to a full state
tomography.
 We  applied our method in systems of dimension
$2\otimes2$, $2\otimes3$ and $3\otimes3$.
Our numerical investigations 
showed that entanglement can be detected with a cost which is much lower
than full state tomography, when the entanglement is not very small.
For low entangled states, including PPT ones, we presented a method to
construct entanglement witnesses (EW). The EWs have an  error bar
that monotonically diminishes with the increase of information about
the state. Our tests suggest that the  error bar is too big, for
ignoring it, 
the entanglement estimate yielded by the EW is always a lower bound
to the true entanglement. Therefore we believe that a tighter error
bar could be calculated, but we weren't able to prove it yet. 

We also discussed the choice of observables to be measured in 
the laboratory. In particular, we noted that the choice is not
obvious in the case of the $2\otimes3$ system, when one is willing
to measure the smallest set of informationally  complete projectors.
Nevertheless, we offered a  method to construct these minimal 
informationally complete sets, in the case of projective measurements.

The application of our approach to multipartite systems is 
straightforward, at the level of the formalism. 
As a matter of fact, we performed some tests on NPT states of
three qubits, obtaining results similar to the ones we presented
for the bipartite systems.

It would be interesting to investigate how our approach could be 
adapted to study other properties, as the purity of a state, for 
example. In this case, quadratic or higher degree non-linear programs
would be necessary.

ACKNOWLEDGEMENTS: Financial support provided
by the Brazilian agencies  FAPEMIG and  CNPQ. 
We thank  S. P\'adua for the fruitful discussions about
measurements.  ROV is indebted to 
G.G. Silva for her overall support.

\end{document}